\documentclass[a4paper,aps,pra,preprintnumbers]{revtex4}
\usepackage[utf8]{inputenc}
\usepackage[english]{babel}
\usepackage[T1]{fontenc}
\usepackage{amssymb,amsfonts,amsmath,mathtext,enumerate,float,dsfont}
\usepackage{graphics,graphicx,epsfig,epstopdf}
\usepackage{caption}
\usepackage{cmap}
\usepackage{multirow}
\usepackage{indentfirst}
\usepackage[usenames]{color}
\usepackage{amsthm}
\usepackage{xcolor}

\begin{document}

\title{Generation of squeezed Fock states by measurement}

\author{S. B. Korolev$^{1,2}$, E. N. Bashmakova$^{1}$, A. K. Tagantsev $^{3}$, T. Yu. Golubeva$^{1,2}$} 
\affiliation{$^1$ St.Petersburg State University, Universitetskaya nab. 7/9, St.Petersburg, 199034, Russia}
\affiliation{$^2$ Laboratory of Quantum Engineering of Light, South Ural State University, pr. Lenina 76, Chelyabinsk, 454080, Russia}
\affiliation{$^3$ Swiss Federal Institute of Technology (EPFL), School of Engineering, Institute of Materials Science, CH-1015 Lausanne, Switzerland}

\begin{abstract}

The generation of squeezed Fock states by the one or more photon subtraction from a two-mode entangled Gaussian (TMEG) state is theoretically addressed. We showed that an arbitrary order Fock state can be generated this way and we obtained a condition that should be imposed on the parameters of the TMEG state to guaranty such a generation. We called the regime, in which this condition is satisfied, {\it universal solution} regime. We showed that, for the first squeezed Fock state, the above condition is redundant such that the generation of the first squeezed Fock state is still possible by a one photon subtraction from an arbitrary TMEG state. At the same time, the maximum generation probability of the first squeezed Fock state generation corresponds to the universal solution regime. We applied the above results to the description of generation of the squeezed Fock states using a beam splitter and a Controlled-Z operation. We have estimated the parameters of such setups and input squeezed states, which are necessary to obtain squeezed Fock states with the maximum probability.
 
\end{abstract}

\date{\today}
\maketitle

\section{Introduction}

Non-Gaussian states and non-Gaussian operations are  vital elements of the modern quantum computation and quantum information in continuous-variable.
The interest to non-Gaussian states and non-Gaussian operations is motivated by a number of issues.
Specifically, the non-Gaussian states are known for their utilization in quantum metrology problems \cite{PhysRevResearch.3.033250} as well as in view of  reducing  the teleportation error \cite{Asavanant2021,Zinatullin2023,Zinatullin2021}.
In quantum computing, it was shown that to build a universal continuous-variable quantum computer, one must be able to implement at least one non-Gaussian operation \cite{Braunstein_2005,Lloyd_1999}.
Another  important reason for the increasing interest in non-Gaussian states in quantum computing, is a potential use of such states in error correction protocols \cite{Ralph_2003,Hastrup_2022}.

In this context, one encounters a problem that can be called "fidelity issue".
The point is that, in all  error correction protocols \cite{PhysRevA.73.012325, Eaton_2019,Vasconcelos:10, PhysRevA.64.012310,Ralph_2003,Hastrup_2022}, the non-Gaussian states proposed as the  logic basis states cannot be exactly generated in conventional evolution processes.
For this reason, one has to approximate the desired  non-Gaussian states with available or potentially experimentally available states.
As a result, one should take care that the fidelity between the original and the approximation is sufficiently close to one.
In view of  this issue, of special interest would be a non-Gaussian state which, in one side, is efficient in an error correction protocol, and which, on the other side, can be obtained as the result of  a conventional evolution process.
In this case,  the fidelity issue will be lifted such that one may speak about a 100\% fidelity for the non-Gaussian state used in the protocol.

As a good candidate for the aforementioned state, one can consider the squeezed Fock (SF) states \cite{PhysRevLett.119.033602,PhysRevA.87.062115,Olivares2006,olivares2005squeezed,Kral1990,NIETO1997135,PhysRevA.40.2494}. With resects to their performance in error correction protocols, as was recently shown \cite{QECCSFOCK}, SF states  are quite competitive  with the  Schrodinger cat states  \cite{Sychev2017,Buzek1995} and the squeezed Schrodinger cat states \cite{bashmakova2023effect,Ourjoumtsev2007}, which were  proposed for such protocols \cite{Ralph_2003,Hastrup_2022}. Also, SF states are considered as a useful resource for quantum metrology \cite{Zhang, Hou} to improve the phase sensitivity. SF states are used to increase the fidelity of swap operations \cite{DellAnno}. In addition, it has been recently shown \cite{Winnel} that such states are a powerful resource for the generation of other useful non-Gaussian states, such as the large-amplitude squeezed cat and high-quality Gottesman-Kitaev-Preskill (GKP) states. As for a possibility to generate a SF state in  a conventional evolution process, it was shown \cite{Xu2015,olivares2005squeezed} that it is possible by the subtraction of a photon from the state resulted from the interference of the  vacuum and squeezed vacuum at a beam splitter.
This process can be  viewed as the result of  detection of the first Fock state in one mode of a two-mode entangled Gaussian (TMEG) state.
In this context, the following reasonable questions arise.
Specifically, if the above TMEG state is the only one, for which  the  detection of one photon in one mode results in the generation of  the first squeezed  Fock state in the other.
If no, what is the subclass of the TMEG states where such a procedure leads to the generation of the first SF state.
Of interests is also to find out if the detection of a higher-order Fock state in one mode can lead to the generation of a   higher-order SF state in the other. If yes, one is interested in the question of what is the subclass of  the TMEG states where higher-order SF states can be generated by measurement.

This paper is devoted to  answering the questions listed above.
We prove that the detection of  one photon  in one mode of an arbitrary TMEG state should result in the generation of the first SF in the other.
This statement does not hold if  two or more photons are detected.
However, is this case, as we demonstrate, there exist  conditions which should be imposed on a TMEG state to  guarantee the generation of SF states of an arbitrary order.
We also address the probability of  generation of a SF state under the condition of its fixed degree of squeezing.
We present a compact expression for such a probability and treat the problem of its maximization.
As an implementation of our general theory, we present the results of its application to the experimentally available systems, which are based on  a beam splitter (BS) or a controlled-Z (CZ) gate.

\section{Generation of squeezed Fock states}
\label{Gen}
We consider the general non-displaced TMEG state, which has the following wave function \cite{Rendell2005}:

\begin{align}
\label{TMGS}
    \Psi \left(x_1,x_2 \right)=\frac{\left(\mathrm{Re}[ a] \mathrm{Re} [d]-\left(\mathrm{Re }[b] \right)^2\right)^{1/4}}{\sqrt{\pi}}\exp \left[{-\frac{1}{2} \left(a x_1^2+2 b x_1 x_2+d x_2^2\right)}\right],
\end{align}
where  $a,b,d  \in \mathbb{Z}$ satisfy the following conditions
\begin{align}
 \label{param_TMSS}
  \mathrm{ Re}[d] >0, \qquad   \mathrm{ Re}[a] >0, \qquad  \mathrm{Re} [a] \mathrm{Re}[ d]-\left(\mathrm{Re }[b] \right)^2 >0.
\end{align}
Hereafter, all wave functions are written in terms of the eigenvalues of the coordinate quadrature operator defined by the following relation
\begin{align}
 \label{x}
\hat{x}=(\hat{a}+\hat{a}^\dag)/\sqrt{2}
\end{align}
 where $\hat{a}$ stands for the annihilation operator of the quantum field.
Suffixes 1 and 2   specify the two modes.

We are interested in the state generated in mode 2 by the measurement of  number of photons in mode 1.
 The wave function of the measured n-photon Fock state reads \cite{NIETO1997135}
\begin{align}
\label{FS}
\Psi_{\mathrm{\mathrm{F}}}(x, n)=\frac{e^{-\frac{1}{2} x^2} H_n(x)}{\sqrt{A_n}},
\end{align}
where
\begin{align}
    A_n=2^n n! \sqrt{\pi},
\end{align}
and $H_n$ stands for the Hermite polynomials \cite{gradshteyn2014table}   defined such that the three  lowest of them read
\begin{equation}
\label{Her}
H_0(x) = 1, \qquad  H_1(x) = 2x,\qquad \mathrm{and}  \qquad H_2(x) = 4x^2-2.
\end{equation}
Using the von Neumann postulate, the wave function of the state in mode 2, can be found as follows:
 \begin{multline}
  \label{int_calc}
  \Psi_{\mathrm{out}}\left(x_2,n\right)=\frac{1}{\sqrt{P_n}} \int\Psi_{\mathrm{F}}^*(x_1, n)  \Psi \left(x_1,x_2 \right) dx_1\\
  =\frac{\left(\mathrm{Re }[a] \mathrm{Re} [d]-\left(\mathrm{Re} [b] \right)^2\right)^{1/4}}{\sqrt{\pi A_n P_n}}\int e^{-x_1^2/2}H_n(x_1)e^{-\frac{1}{2} \left(a x_1^2+2 b x_1 x_2+d x_2^2\right)}dx_1,
\end{multline}
where $P_n$ is the probability of measuring  $n$ photon in mode 1. To obtain an explicit expression for the output wave function, one should evaluate the following integral
\begin{align}
\label{I1}
       I_n=\int H_n(x_1)e^{-\frac{1}{2} \left(a+1\right) x_1^2-b x_1 x_2}dx_1.
   \end{align}
Using the definition of generating function of Hermite polynomials  \cite{Babusci2012}
 \begin{align}
  \label{A_epf}
 e^{2x_1t - t^2} &= \sum_{n=0}^\infty H_n\left[x_1\right] \frac{t^n}{n!}
\end{align}
we can write
\begin{align}
\begin{aligned}
\\&\sum_{n=0}^\infty \frac{t^n}{n!} I_n=\int _{-\infty}^{\infty}e^{2x_1t - t^2}e^{-\frac{1}{2} \left(a+1\right) x_1^2-b x_1 x_2}dx_1
\\&
 =\sqrt{\frac{2\pi}{a+1}}\exp \left[-\frac{2bx_2}{a+1}t-\left(\frac{a-1}{a+1}\right)t^2+\frac{b^2x_2^2}{2\left(a+1\right)}\right]
 \\&=\sqrt{\frac{2\pi}{a+1}}e^{\frac{b^2x_2^2}{2\left(a+1\right)}}\sum_{n=0}^\infty H_n\left[-\frac{bx_2 \sqrt{a+1}}{(a+1)\sqrt{a-1}}\right] \left(\frac{a-1}{a+1}\right)^{n/2}\frac{t^n}{n!}.
 \label{I2}
\end{aligned}
 \end{align}
Next Eq.(\ref{I2}) implies:
\begin{align}
\label{I3}
I_n=(-1)^n\sqrt{\frac{2\pi}{a+1}}e^{\frac{b^2x_2^2}{2\left(a+1\right)}}
H_n\left[\frac{bx_2}{\sqrt{a^2-1}}\right]  \left(\frac{a-1}{a+1}\right)^{n/2}.
 \end{align}
Using this relationship, the function $\Psi_{\mathrm{out}}\left(x_2,n\right)$ can be expressed in terms of polynomials $H_n(x)$:

\begin{align} \label{Psi_out}
\Psi_{out}\left(x, n\right)=
\frac{(-1)^n \sqrt[4]{\mathrm{Re} [a] \mathrm{Re} [d]-\left(\mathrm{Re }[b] \right)^2}}{\sqrt{A_n P_n} }
\sqrt{\frac{2(a-1)^n}{(a+1)^{n+1}}}
e^{-\frac{1}{2} x^2
\left(d-\frac{b^2}{a+1}\right)}
 H_n\left[\frac{b x}{\sqrt{a^2-1}}\right].
\end{align}

Now we are looking for conditions which provide the generation of the $n^{th}$ SF state, having a wave function:
\begin{align}
\label{SFS}
\Psi_{\mathrm{\mathrm{SF}}}(x, n)=\frac{e^{-\frac{1}{2} e^{2 r} x^2} H_n\left(e^r x\right)}{\sqrt{A_n e^{-r}}},
\end{align}
once the $n$ photons are detected. Here, $r$ is the real squeezing parameter, which can take both positive and negative values; positive and negative $r$ correspond to the $\hat{x}$-quadrature and $\hat{y}$-quadrature squeezing, respectively.

Comparing directly Eqs. (\ref{Psi_out}) and (\ref{SFS}), we conclude that the SF state with a given  $r$ is generated if
\begin{align}
    &d-\frac{b^2}{a+1}=e^{2r}, \label{cond0}\\
    &\frac{b^2}{a^2-1}=e^{2r} \label{cond1}.
\end{align}
These conditions are obtained from the equality of exponentials and polynomials for two wave functions (\ref{Psi_out}) and (\ref{SFS}). Since both wave functions are normalized, the equality of their functional dependencies implies the complete equality of the two wave functions (up to a common phase). To satisfy  (\ref{cond0}), one should also require that

\begin{align} \label{Re}
    \mathrm{Re}\left[d-\frac{b^2}{a+1}\right]>0
\end{align}
and
\begin{align} \label{Im}
    \mathrm{Im}\left[d-\frac{b^2}{a+1}\right]= 0.
\end{align}
Here, (\ref{Re}) directly follows from (\ref{param_TMSS}), as one can readily check.

Thus, Eqs.  (\ref{cond0}), (\ref{cond1}), and (\ref{Im}) yield the conditions that should be imposed  on the parameters of the TMEG state to guarantee to generate  the $n^{th}$ SF state in one mode once  the n-photon state is detected in the other.

The above consideration was essentially based on condition (\ref{Im}).
If we lift this restriction, starting from (\ref{int_calc}) one can derive the following  output wave function
\begin{align}
\label{SFSR}
\Psi_{\mathrm{\mathrm{SFR}}}(x, n)=\frac{e^{-\frac{1}{2}(e^{2r_\theta}+iw)x^2} H_n(e^{r_\theta}x)}{\sqrt{e^{-r_\theta}A_n}}.
\end{align}
where
\begin{align}
 e^{2r_\theta}= \mathrm{Re}\left[d-\frac{b^2}{a+1}\right], \qquad w=\mathrm{Im}\left[d-\frac{b^2}{a+1}\right],
 \label{cond2}
\end{align}
\begin{align}
 e^{2r_\theta}=\frac{b^2}{a^2-1}.
 \label{cond3}
\end{align}
Such a form of the wave function corresponds to the SF state which is rotated in the quadratures' plane by angle $\theta$ that meets the following equation
\begin{align}
\label{angle}
\text{tan}(2\theta)=\frac{2
   w}{e^{4 r_\theta}+w^2-1}.
\end{align}
This can be shown basing on the results from Ref.\cite{NIETO1997135}.
Actually,  such a rotation eliminates the imaginary contribution to the exponent in (\ref{SFSR}) bringing it to the form given by Eq.(\ref{SFS}),  however, with the following squeezing factor
\begin{align}
\label{squeez}
\cosh r=\frac{e^{-r_{\theta }}}{2}  \sqrt{2 e^{2 r_{\theta }}+e^{4 r_{\theta }}+w^2+1}.
\end{align}
The explicit relationships between the scheme parameters that guarantee  the generation of the rotated SF state are given in Appendix \ref{append_RSF}.

Hereafter, we will focus on the generation of the non-rotated SF states.

In the above consideration, we were looking for a condition that  should be imposed on the values of parameters of  a TMEG state to provide the generation of the $n^{th}$ SF state in one mode once  $n$ photons are detected in the other mode. Such a condition reads
\begin{align} \label{condF}
\mathrm{Re}\left[d-\frac{b^2}{a+1}\right]=\frac{b^2}{a^2-1}.
\end{align}
This condition follows from the combination of Eqs. (\ref{cond0}), (\ref{cond1}) and the fact that inequality (\ref{Re}) is always satisfied. Hereafter, the situation, where (\ref{condF}) is met, we term \textit{universal solution regime}.

At this point, one notices  that the case where $n=1$ is special.
Namely,  it occurs that  condition  (\ref{condF}) is required  only for $n\geq 2$ while for $n=1$ it is redundant. This is seen from Eq. (\ref{Psi_out}) if one takes into account  that,  for $n=1$,  $H_1(x)$ is a monomial. The implication is  that the one-photon subtraction from an arbitrary  TMEG state will always result in the generation of the first SF state. This statement holds for the case of the aforementioned rotated SF states either.
\section{Probability of generation}
 \label{general}
In the previous section, we addressed the  generation of SF states with a given squeezing. Now we examine the probabilities of such a generation.
Let us  first find  the probability, $P_n$,  of  measuring n photons in one mode of an arbitrary  TMEG state.
Using  Born's rule
 \begin{equation}
 \label{Born}
P_n=\int \left|\int\Psi_{\mathrm{F}}^*(x_1, n)  \Psi \left(x_1,x_2 \right) dx_1 \right|^2dx_2
 \end{equation}
and Eqs.(\ref{TMGS}) and (\ref{FS}), we arrive at the result  expressed (see Appendix  \ref{appendix_norm}) in terms of the hypergeometric  function $ _2F_1(x,y;s;t)$ \cite{gradshteyn2014table}:
\begin{align}
 \label{norm}
    P_n=
   \frac{2| b| ^{2n} \sqrt{\mathrm{Re}[ a] \mathrm{Re} [d]-\left(\mathrm{Re}[ b] \right)^2}}{\left(|1+a| ^2 \mathrm{Re} \left[d-\frac{b^2}{a+1}\right]\right)^{n+1/2}} \,
   _2F_1\left[\frac{1-n}{2},-\frac{n}{2};1;\left| 1-\frac{\left(a^2-1\right) \mathrm{Re} \left[d-\frac{b^2}{a+1}\right]}{b^2}\right| ^2\right].
\end{align}

We are interested in the probability of measuring n photons in one mode of a TMEG state, the parameters of which correspond to the   generation  a SF state with a given squeezing degree in the other mode.

First, we address the universal solution regime where  conditions (\ref{cond0}),  (\ref{cond1}), and (\ref{Im})  make possible  the generation of SF states  for any $n$.
In this case,  $\left(a^2-1\right) \mathrm{Re} \left[d-\frac{b^2}{a+1}\right]=b^2$ such that the hypergeometric  function entering Eq.(\ref{norm}) reads
\begin{align}
 \label{F}
\, _2F_1\left(\frac{1-n}{2},-\frac{n}{2};1;0\right)=1.
\end{align}
while Eq. (\ref{norm}) becomes:
\begin{align}
 \label{pr}
P_n=\frac{2 \left|a^2-1\right| ^n\sqrt{\mathrm{Re}[a]^2-\mathrm{Re}\left[\sqrt{a^2-1}\right]^2}]}{|1+ a|^{2n+1}}.
\end{align}

One readily notices that (i)  the probability is   independent of squeezing of the generated state, and (ii) it is a function of a free parameter, $a$, such that it can be maximized holding the squeezing.
For  a SF state with a certain number $n$, the maximization of (\ref{pr}) yields the maximum probability
\begin{align}
 \label{prma}
    P_{n, \mathrm{max}}=\frac{n^n}{(1+n)^{n+1}},
\end{align}
which is reached at  $a$ satisfying the following condition:
\begin{align} \label{rings}
    \left(a''\right)^2+\left(a'-\left(n+\frac{1}{4 n+2}+\frac{1}{2}\right)\right)^2=\frac{4 n^2 (n+1)^2}{(2 n+1)^2}
\end{align}
where $a'$ and $a''$ are the real and imaginary parts of the parameter $a=a'+ia''$.
Equation (\ref{rings}) defines a set of circles on the complex plane $a$.

One readily checks that,  for real $a$,  two solutions to (\ref{rings}) read
 \begin{align}
  \label{am1}
    a_{1,\text{n}}=\frac{1}{2n+1} ,
\end{align}
 \begin{align}
  \label{am2}
  a_{2,\text{n}}=1+2n.
\end{align}

Outside the universal solution regime addressed above, as was shown in Sect. \ref{Gen}    the generation  the SF state is possible  only for $n=1$.
Being interested in the non-rotated  first SF state, the condition (\ref{Im}) is  met.
Thus, for  $n=1$,  the probability of generation of the first SF state  given by Eq. (\ref{norm}) reads:
 \begin{align}
  \label{norm1}
    P_1=\frac{2| b| ^{2} \sqrt{\mathrm{Re}[ a] \mathrm{Re} [d]-\left(\mathrm{Re}[ b] \right)^2}}{\left(|1+a| ^2 e^{2r}\right)^{3/2}}
\end{align}
where
\begin{align}
 \label{oneregime}
   b^2=(a+1) \left(d-e^{2 r}\right).
\end{align}

The latter follows form (\ref{cond0}). Thus, we see that now, in contrast to the  universal solution,  (i)  the probability is dependent on the squeezing of the generated state, and (ii) it is a function of two free parameters: $a$ and $d$. Here one might expect that  the presence of the two free parameters instead one will provide a more efficient maximization of the probability.
However, it occurs that this is not the case.  Estimating the extremum of Eq. (\ref{norm1}), we find that the maximum of this function is $0.25$, which is reached once the values of parameters of TMEG state correspond to the  universal solution. Thus, for the case of generation of the first SF state, the universal solution corresponds to the maximum probability.

\section{Implementation of the general theory}
 \label{sec_pract_imp}

 The above sections addressed the problem of generation of a SF state starting from of a TMEG state. Now the results obtained are  implemented in the description of such a generation for two situations where the TMEG state  appears in customary experimental setups.

\begin{figure}
    \centering
    \includegraphics[scale=0.85]{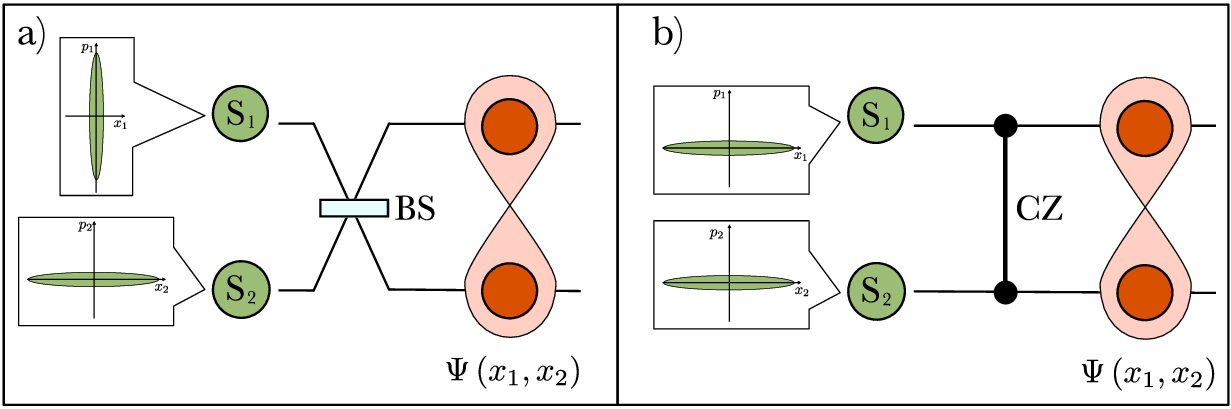}
    \caption{Schemes for generating a  TMEG state: a) with real coefficients $a$, $b$ and $d$ b) with real coefficients $a$, $d$ and imaginary $b$. In the figure: $\text{S}_i$ is the i-th squeezed oscillator, BS is the beam splitter, CZ is the CZ transformation, $\Psi \left(x_1,x_2\right)$ is the wave function of the output TMEG state. In  inserts,  the cross-sections of the Wigner function over the quadrature plane are schematically depicted to illustrate the squeezing of the input. }
    \label{fig:BS_CZ_gen}
\end{figure}

  The setups of  interest are shown in Fig. 1. First setup (Fig. \ref{fig:BS_CZ_gen} a), in which two orthogonally squeezed vacuum states  are interfered at a BS \cite{Takase2021,bashmakova2023effect}. In another experimental setup (Fig. \ref{fig:BS_CZ_gen} b), two squeezed vacuum states with the parallel orientation of squeezing  pass through a CZ gate \cite{PhysRevA.103.062407,PhysRevA.98.042304,PhysRevA.102.042608}.
 We will consider the situation where the degree of squeezing of the two inputs,  $|r_1|$ and $|r_2|$, are not necessarily equal,  ${r_1}$ and ${r_2}$ being the corresponding real squeezing parameters. Let us note that the real squeezing parameter can take both positive and negative values.  For the BS setup, ${r_1}$, ${r_2}$,  and the  power transmission  of BS $t$ can be related with the parameters of the starting  TMEG state by the following expressions  \cite{Takase2021}:
\begin{align}
   &a= e^{2 r_1} (1-t)+e^{2 r_2}t, \label{param_a}\\
   &b=\sqrt{(1-t) t} \left(e^{2 r_1}-e^{2 r_2}\right), \label{param_b}\\
    &d=e^{2 r_1} t+e^{2 r_2} (1-t) \label{param_d},
\end{align}
where the real squeezing parameters satisfy the condition: $r_1r_2<0$. This requirement follows from the fact that two orthogonally squeezed states are interfered at BS. For the CZ setup, one should relate the parameters of the starting  TMEG state with ${r_1}$,  ${r_2}$, and the real weight coefficient $g$, which parameterizes  the idealized  operator of the CZ transformation \cite{PhysRevA.103.062407,PhysRevA.98.042304,PhysRevA.102.042608}
\begin{align} \label{CZ}
    \Hat{U}_{CZ}=e^{ig \hat{x}_1\hat{x}_2}
\end{align}
where $\hat{x}_i$ is the quadrature operator of the i-th oscillator (see Eq. (\ref{x})).
One can show that \cite{PhysRevA.103.062407,PhysRevA.98.042304,PhysRevA.102.042608}
\begin{align}
    &a=e^{2r_1},  \label{CZ1}\\
    &b=ig, \label{CZ2}\\
    &d=e^{2r_2}. \label{CZ3}
\end{align}
One readily checks that in both cases parameters  $a$  and $d$ are real.
As for $b$, for the BS case, it  is real while, for the CZ case,  it is imaginary.

We are interested in a description of both BS and CZ cases, in the universal solution regime and outside it.
For the universal solution regime, using (\ref{cond1}) and (\ref{condF}), for  the BS case, since $b$ is real, we find that  $a>1$  while,  for  the CZ case, since $b$ is imaginary, we find that  $a<1$.
As a result, the following relations can be written
\begin{align}
 \label{BS}
    b=\pm\sqrt{a^2-1}e^{r}, \qquad  d=a e^{2r}
\end{align}
and
\begin{align}
 \label{CZ0}
    b=\pm i\sqrt{1-a^2}e^{r}, \qquad d=a e^{2r}.
\end{align}
for the BS and  CZ case,  respectively.
These relations imply that, in  the universal solution regime, the parameters $r$ and  $a$, the values of which can be arbitrary chosen, fully define  the required TMEG state for the generation of the $n$ SF state with an arbitrary $n$. Outside that regime, where the generation of the first SF state is only possible,  parameters $r$,  $a$,  and $d$, the values of which can be arbitrary chosen, fully define the required TMEG state, with $b$ being given by (\ref{oneregime}).

Now, let us focus on the BS case. In the universal solution regime,  for a given squeezing of the output SF state, parameter $a$ controls the situation. Specifically, with a fixed $a$, by solving the set of equations (\ref{BS}),  (\ref{param_a}), (\ref{param_b}), and (\ref{param_d}), one can find the power transmission of BS $t$  and the squeezing parameters of the inputs $r_1$ and $r_2$, which are required to generate an arbitrary SF state:
\begin{align}
    &r_1+r_2=r \label{rout}\\
    &t=\frac{1}{1-\frac{\sinh r_2 \cosh r_2}{\sinh r_1 \cosh {r_1}}}, \quad \text{for} \quad r_1r_2<0.
\end{align}
Equation (\ref{rout}) reveals  a remarkable feature of the universal solution regime: in this regime, for the BS case, the squeezing of the  output SF state is just the sum of squeezing of the inputs.  This equation is important for practical applications giving the squeezing parameter of the resultant SF state as a function of squeezing of the input.

As for the probability of the generation, it can be found using (\ref{pr}).
If one is  interested in the situation with the maximum generation probability,  the value of  $a$ maximizing the probability is given by Eq. (\ref{am2}). This imposes an additional condition on the squeezing parameters of the input states: 
\begin{align}
    &r_1=\frac{1}{2} \ln \left(1+2n+\frac{4 (n+1) n}{e^{-2
   r_2}-1-2n}\right), \label{rimp}\\
   &|r_2|>\frac{1}{2}\ln \left(1+2 n\right).
\end{align}

Outside the universal solution regime,  for a given squeezing of the output SF state, the situation is controlled parameters $a$ and  $d$.
Now the required  values of the experimental parameters can be found by solving the set of equations (\ref{oneregime}),  (\ref{param_a}), (\ref{param_b}), and (\ref{param_d}) while the probability of the generation is given by (\ref{norm1}).

As an illustration for the calculation procedure outlined above, in Table \ref{tab:BS},  we present some examples of values of the experimental parameters required for the generation of some SF states using a BS.
Columns  "Maximized probability case"\, and "Non-maximized probability case"\, are dealing with the universal solution regime for  values of $a$ which maximize the  probability (see Eq. (\ref{am2})) and for  values of $a$ which do not. Columns  "Vacuum in one channel"\, show some  results for the situation which lays outside the universal solution regime. It is the situation treated earlier \cite{olivares2005squeezed,Podoshvedov_2023} where one of the inputs is not squeezed. For these columns, the values of the parameters given in the Table corresponds to the maximized  probability. The fact that the case where the vacuum in one channel lies outside the universal solution means that in this scheme, we can only generate the first SF state. For the cases presented in this Table, the following  trend may be seen: for a fixed squeezing of the output SF state,  by scarifying the generation probability one can reduce the degree of squeezing of the  input.
\begin{table}[h!]
\begin{tabular}{|l|ll|ll|ll|}
\hline
                     & \multicolumn{2}{c|}{Maximized probability case}                & \multicolumn{2}{c|}{Non-maximized probability case}  & \multicolumn{2}{c|}{vacuum in one channel}           \\ \cline{2-7} 
                     & \multicolumn{1}{l|}{$r=\frac{1}{2}$ (4.3 dB)} & $r=1$ (8.7 dB) & \multicolumn{1}{l|}{$r=\frac{1}{2}$ (4.3 dB)} & $r=1$ (8.7 dB) & \multicolumn{1}{l|}{$r=\frac{1}{2}$ (4.3 dB)} & $r=1$ (8.7 dB) \\ \hline
\multirow{4}{*}{n=1} & \multicolumn{1}{l|}{$r_1=1.19$ (10.3 dB)}               &     $r_1=1.60$ (14.0 dB)          & \multicolumn{1}{l|}{$r_1=0.96$ (8.5 dB)}             &     $r_1=1.39$ (12.1 dB)            & \multicolumn{1}{l|}{$r_1=0$ (0 dB)}              & $r_1=0$ (0 dB)             \\
                     & \multicolumn{1}{l|}{$r_2=-0.69$ (-6.0 dB)}               &       $r_2=-0.60$ (-5.3 dB)          & \multicolumn{1}{l|}{$r_2=-0.46$ (-4.0 dB)}               &    $r_2=-0.39$ (-3.4 dB)            & \multicolumn{1}{l|}{$r_2=1.02$ (8.9 dB)}               &  $r_2=1.19$ (10.4 dB)               \\
                     & \multicolumn{1}{l|}{$t=0.74$}                      &      $t=0.88$          & \multicolumn{1}{l|}{$t=0.75$}               &     $t=0.90$           & \multicolumn{1}{l|}{$t=0.67$}               &         $t=0.44$       \\
                     & \multicolumn{1}{l|}{$P=25 \%$}                  &    $P=25 \%$            & \multicolumn{1}{l|}{$P=22 \%$}               &      $P=22 \%$          & \multicolumn{1}{l|}{$P=9 \%$}               &    $P=13.5 \%$            \\ \hline
\multirow{4}{*}{n=2} & \multicolumn{1}{l|}{$r_1=1.45$ (12.6 dB)}               &   $r_1=1.86$ (16.2 dB)             & \multicolumn{1}{l|}{$r_1=1.19$ (10.3 dB)}               &        $r_1=1.60$ (14.0 dB)        & \multicolumn{1}{l|}{-NON-}               &       -NON-         \\
                     & \multicolumn{1}{l|}{$r_2=-0.95$ (-8.3 dB)}               &     $r_2=-0.86$ (-7.5 dB)           & \multicolumn{1}{l|}{$r_2=-0.69$ (-6.0 dB)}               &       $r_2=-0.60$ (-5.3 dB)         & \multicolumn{1}{l|}{}               &                \\
                     & \multicolumn{1}{l|}{$t=0.73$}                      &    $t=0.88$            & \multicolumn{1}{l|}{$t=0.74$}               &     $t=0.89$           & \multicolumn{1}{l|}{}               &                \\
                     & \multicolumn{1}{l|}{$P=14.8 \%$}                  &     $P=14.8 \%$           & \multicolumn{1}{l|}{$P=12.5 \%$}               &      $P=12.5 \%$          & \multicolumn{1}{l|}{}               &                \\ \hline
\end{tabular}
\caption{Values of  experimental parameters required for the generation of $n$-th SF states characterized by the squeezing parameters  $r=1/2$ and $r=1$  using a BS. Situations addressed: the maximized generation probability for the universal solution regime with the squeezed vacuum in both input channels, a non-maximized generation probability for the universal solution with the squeezed vacuum in both input channels and  the maximized generation probability for the regime where the  vacuum  is squeezed only in one input channel. In the table $r_i$ are the squeezing coefficients of the two input channels, the brackets show the squeezing coefficients in dB, which are related to $r_{i}$, as $10\,\mathrm{log}_{10} e^{2r_{i}}$} and $t$ is the value of the power transmission of the BS, and $P$ is the probability of generating a SF state for the indicated values of  the experimental parameters.
 \label{tab:BS}
\end{table}

In the CZ case, in the universal solution regime,  for a given squeezing of the output SF state, as in the BS case, parameter  $a$ controls the situation.
Now, for a given squeezing of the output, the required values of  the experimental parameters can be found using Eqs. (\ref{CZ0}),  (\ref{CZ1}),  (\ref{CZ2}), and  (\ref{CZ3}). The explicit relation between the experimental parameters and the parameters of the generated SF state has the following form:
\begin{align}
    &r_2-r_1=r,\label{rimp}\\
    &g=\pm e^r \sqrt{1-e^{4r_1}}, \quad \text{for} \quad r_1<0.
\end{align}
Equation (\ref{rimp}) reveals  a remarkable feature of the universal solution regime for the CZ case: in this regime, the squeezing of the  output SF state equals  the difference between  the squeezing of the inputs. This equation is important for practical applications giving the squeezing parameter of the resultant SF state as a function of squeezing of the input.

In order to generate the SF state with the highest possible probability in the scheme with CZ gate, it is necessary to impose an additional condition:
\begin{align}
    r_1=-\frac{1}{2} \ln (1+2 n).
\end{align}

In the CZ case, outside the universal solution regime,  one should use Eqs.  (\ref{oneregime}),  (\ref{CZ1}),  (\ref{CZ2}), and  (\ref{CZ3}).

As illustration for such a calculation procedure, in Table \ref{tab:CZ},  we present some examples of values of the experimental parameters required for the generation of some SF states using a CZ gate. We restrict ourselves to the universal solution regime, considering the values of $a$ which maximize the  probability (see Eq. (\ref{am1}))  and the  values of $a$ which do not.

\begin{table}[h!]
\begin{tabular}{|l|ll|ll|}
\hline
                     & \multicolumn{2}{c|}{Maximized probability case}                      & \multicolumn{2}{c|}{Non-maximized probability case}                  \\ \cline{2-5} 
                     & \multicolumn{1}{l|}{$r=\frac{1}{2}$ (4.3 dB)} & $r=1$ (8.7 dB)       & \multicolumn{1}{l|}{$r=\frac{1}{2}$ (4.3 dB)} & $r=1$ (8.7 dB)       \\ \hline
\multirow{4}{*}{n=1} & \multicolumn{1}{l|}{$r_1=-0.55$ (-4.8 dB)}     & $r_1=-0.55$ (-4.8 dB) & \multicolumn{1}{l|}{$r_1=-0.80$ (-7.0 dB)}     & $r_1=-0.80$ (-7.0 dB) \\
                     & \multicolumn{1}{l|}{$r_2=-0.05$ (-0.4 dB)}     & $r_2=0.45$ (3.9 dB)  & \multicolumn{1}{l|}{$r_2=-0.30$ (-2.7 dB)}     & $r_2=0.20$ (1.6 dB) \\
                     & \multicolumn{1}{l|}{$g=1.55$}                 & $g=2.56$             & \multicolumn{1}{l|}{$g=1.62$}                 & $g=2.66$             \\
                     & \multicolumn{1}{l|}{$P=25 \%$}                & $P=25 \%$            & \multicolumn{1}{l|}{$P=22 \%$}                & $P=22 \%$            \\ \hline
\multirow{4}{*}{n=2} & \multicolumn{1}{l|}{$r_1=-0.80$ (-7.0 dB)}     & $r_1=-0.80$ (-7.0 dB) & \multicolumn{1}{l|}{$r_1=-0.55$ (-4.8 dB)}     & $r_1=-0.55$ (-4.8 dB) \\
                     & \multicolumn{1}{l|}{$r_2=-0.30$ (-2.6 dB)}     & $r_2=0.20$ (1.7 dB) & \multicolumn{1}{l|}{$r_2=-0.05$ (-0.4 dB)}     & $r_2=0.45$ (3.9 dB) \\
                     & \multicolumn{1}{l|}{$g=1.62$}                 & $g=2.66$             & \multicolumn{1}{l|}{$g=1.55$}                 & $g=2.56$             \\
                     & \multicolumn{1}{l|}{$P=14.8 \%$}              & $P=14.8 \%$          & \multicolumn{1}{l|}{$P=12.5 \%$}              & $P=12.5 \%$          \\ \hline
\end{tabular}
 \caption{
 Values of  experimental parameters required for the generation of $n$-th SF states characterized by the squeezing parameters  $r=1/2$ and $r=1$  using the CZ transformation.
 Situations addressed: the maximized generation probability for the universal solution regime with the squeezed vacuum in both input channels and a non-maximized generation probability for the universal solution with the squeezed vacuum in both input channels. In the table $r_i$ are the squeezing coefficients of the two input channels, the brackets show the squeezing coefficients in dB, which are related to $r_{i}$, as $10\,\mathrm{log}_{10} e^{2r_{i}}$} and $g$ is the value of the weight coefficient of the CZ transformation, and $P$ is the probability of generating a SF state for the indicated values of  the experimental parameters.
 \label{tab:CZ}
\end{table}

\section{Conclusion}

In this work, we  addressed the problem of generation of squeezed Fock (SF) states by one or more photon subtraction  from a non-displaced two-mode entangled Gaussian (TMEG) state.
We demonstrated that  no matter what a TMEG state is used, in the one-photon-substation setup, one will always get a SF state in the output.
For a given squeezing of the generated state, the body of the suitable states can be parameterized with two free parameters.
For a two or more photon  subtraction, an arbitrary TMEG state is not suitable for the  generation SF states.
However, we showed that, in this  case, there exists a body of TMEG state, which provides the generation of higher-order SF states.
We derived the conditions, which should be imposed on the parameters of the TMEG state, to make the  generation possible.
We termed  the regime where such conditions are met universal solution regime.
In this regime, for a given squeezing of the generated state, one  free parameter controls the situation.
We derived compact expressions for generation probability for the universal solution regime and outside of such a regime.
For the universal solution regime, where the $n^{th}$ SF state can be generated  for an arbitrary $n$, the generation probability was found to be independent of the squeezing degree of the generated state.
Outside the universal solution regime, where a SF state can be generated only for $n=1$, the generation probability was found to be dependent on the squeezing degree of the generated state.
Remarkably, it was found that,  for $n=1$,  the generation probability is maximum for the universal solution regime.

As an application of the above general theory we addressed the generation of SF states in the case of  two common experimental setups used  for the generation of  TMEG states: (i)
a beam splitter (BS) fed with   two orthogonally squeezed vacuum states  and (ii) a setup where two squeezed vacuum states with the parallel orientation of squeezing  pass through a CZ gate.
We presented relationships which enables one to determine the values of parameters of the setups (the squeezing at the inputs and the BS power  transmission for BS and the weight coefficient for CZ), which are requited to generate a specified SF state. We identified a remarkable feature of the universal solution regime. In this regime,  the squeezing of the output SF state, for the BS case, is  the sum of squeezing of the inputs while, for the CZ case, it is the difference. For a number of selected situations, the numerical values of parameters of the setups, which are required for the generation of  a specified SF state, were provided.

\section*{Funding}
This research was supported by the Theoretical Physics and Mathematics Advancement Foundation "BASIS" (Grants No. 21-1-4-39-1). SBK and TYG acknowledge support by the Ministry of Science and Higher Education of the Russian Federation on the basis of the FSAEIHE SUSU (NRU) (Agreement No. 075-15- 2022-1116).

\section*{Disclosures}
The authors declare no conflicts of interest.

\section*{Data availability} Data underlying the results presented in this paper are not publicly available at this time, but may be obtained from the authors upon reasonable request.

\appendix

\section{Rotated SF state generation} \label{append_RSF}
We are interested in the generation of a rotated SF state that have the following vector of state:
\begin{align} \label{A_1}
    |\xi,n\rangle =\hat{S}\left(\xi\right)|n\rangle,
\end{align}
where $\hat{S}\left(\xi\right)$ is the squeezing operator, $\xi=|r| e^{i\varphi}$ is the squeezing parameter and  $|n\rangle$ is the vector of state of the n-th Fock state. The  vector of state (\ref{A_1}) has the following wave function in the coordinate representation \cite{NIETO1997135}:
\begin{align} \label{RSFS}
    \Psi_{\mathrm{SFR}}\left(x,n\right)=\frac{\exp \left(\frac{-x^2 (1+i \sin \varphi \sinh 2|r|)}{2 (\cosh 2 |r|-\cos \varphi \sinh 2 |r|)}\right) H_n\left(\frac{x}{\sqrt{\cosh 2|r|-\cos \varphi \sinh 2 |r|}}\right)}{\sqrt[4]{\pi}\sqrt{2^n n!} \sqrt[4]{  \cosh 2 |r|-\cos \varphi \sinh 2 |r|}}.
\end{align}
Comparing Eqs. (\ref{Psi_out}) and (\ref{RSFS}), we can relate the parameters of the TMEG state to those of the rotated SF state by the following relationships:
\begin{align}
 &b=\frac{\sqrt{a^2-1}}{\sqrt{\cosh 2 |r|-\sinh 2|r| \cos \varphi}},\\
 &d=\frac{a+i \sinh 2 |r| \sin \varphi }{\cosh 2 |r|-\sinh 2 |r| \cos \varphi}.
\end{align}

\section{Normalization} \label{appendix_norm}
For a given TMEG state,  the probability to measure n photons in one mode reads:
 \begin{align}
P_n=\int \left|\int\Psi_{\mathrm{F}}^*(x_1, n)  \Psi \left(x_1,x_2 \right) dx_1 \right|^2dx_2.
\end{align}
Starting from Eqs. (\ref{TMGS})  and (\ref{FS}) one finds:
\begin{align} \label{B_2}
P_n=\frac{2\sqrt{\mathrm{Re} [a] \mathrm{Re} [d]-\left(\mathrm{Re }[b] \right)^2}}{A_n}\frac{\left|a-1\right|^n}{\left|a+1\right|^{n+1}}\int e^{-x^2 \mathrm{Re}\left[ d-\frac{b^2}{a+1}\right]}
 H_n\left[\frac{b x}{\sqrt{a^2-1}}\right] H_n\left[\frac{b^* x}{\sqrt{(a^*)^{2}-1}}\right] dx.
\end{align}
Using the results of the work of \cite{Babusci2012}, the integral entering (\ref{B_2}) can be rewritten in the form:
\begin{align}
 &I_n\equiv\int e^{-x^2 \mathrm{Re}\left[ d-\frac{b^2}{a+1}\right]}
 H_n\left[\frac{b x}{\sqrt{a^2-1}}\right] H_n\left[\frac{b^* x}{\sqrt{(a^*)^{2}-1}}\right] dx\\
 &=\frac{1}{2}\sqrt{\frac{\pi}{\alpha }}\sum_{k=0}^{n} \frac{(n!)^2 \left((-1)^{n-k}+1\right) \left(\frac{2\left|\beta\right|^2}{\alpha }\right)^k \left(\frac{\left|\alpha -\beta ^2\right|^2}{\alpha ^2}\right)^{\frac{n-k}{2}}}{ k! \left(\left(\frac{n-k}{2}\right)!\right)^2}
\end{align}
where the following abbreviations have been introduced: $\alpha=\mathrm{Re}\left[ d-\frac{b^2}{a+1}\right]$ and $\beta=\frac{b}{\sqrt{a^{2}-1}}$. The presence of the factor $\left((-1)^{n-k}+1\right)$ means that the sum contains only terms whose numbers satisfy the relation $n-k=2l$. Taking this into account, we can rewrite the sum as:
\begin{align}
   I_n=\sqrt{\frac{\pi}{\alpha }}\sum_{l=0}^{n} \frac{ (n!)^2\left(\frac{2\left|\beta\right|^2}{\alpha }\right)^{n-2l} \left(\frac{\left|\alpha -\beta ^2\right|^2}{\alpha ^2}\right)^{l}}{(n-2l)! \left(l!\right)^2}=n!\left(\frac{2\left|\beta\right|^2}{\alpha }\right)^{n}\sqrt{\frac{\pi}{\alpha }}\sum_{l=0}^{n} \frac{ n! \left(\frac{\left|\alpha -\beta ^2\right|^2}{4\left|\beta\right|^4}\right)^{l}}{(n-2l)! \left(l!\right)^2}.
\end{align}
For brevity, the sum on the right-hand side can be written as a hypergeometric function \cite{gradshteyn2014table}:
\begin{align}
   \sum_{l=0}^{n} \frac{ n! \left(\frac{\left|\alpha -\beta ^2\right|^2}{4\left|\beta\right|^4}\right)^{l}}{(n-2l)! \left(l!\right)^2}= \, _2F_1\left(\frac{1-n}{2},-\frac{n}{2};1;\frac{\left|\alpha -\beta ^2\right|^2}{\left|\beta\right|^4}\right).
\end{align}
Combining all the results, we get the final expression for the probability:
\begin{align}
    P_n=
   \frac{2| b| ^{2n} \sqrt{\mathrm{Re}[ a] \mathrm{Re} [d]-\left(\mathrm{Re}[ b] \right)^2}}{\left(|1+a| ^2 \mathrm{Re} \left[d-\frac{b^2}{a+1}\right]\right)^{n+1/2}} \,
   _2F_1\left[\frac{1-n}{2},-\frac{n}{2};1;\left| 1-\frac{\left(a^2-1\right) \mathrm{Re} \left[d-\frac{b^2}{a+1}\right]}{b^2}\right| ^2\right].
\end{align}

\end{document}